\documentclass[12pt]{spieman}  
\usepackage{amsmath,amsfonts,amssymb}
\usepackage{graphicx}
\usepackage{setspace}
\usepackage{tocloft}
\usepackage{wrapfig}
\usepackage{makecell}
\usepackage{mathtools}
\usepackage[dvipsnames]{xcolor}
\usepackage{lineno}

\DeclarePairedDelimiter\abs{\lvert}{\rvert}
\makeatletter 
\let\oldabs\abs
\def\abs{\@ifstar{\oldabs}{\oldabs*}} 
\makeatother 

\title{Double-offset Cassegrain telescopes for the Ultraviolet Type Ia (UVIa) mission concept}

\author[a,b,*]{Fernando Cruz Aguirre}
\author[b,a]{Keri Hoadley}
\author[c]{Curtis McCully}
\author[d]{Gillian Kyne}
\author[d]{Shouleh Nikzad}
\author[d]{John Hennessy}
\author[d]{April D. Jewell}
\author[d]{Christophe Basset}
\author[c]{Daniel Harbeck}
\author[b,a]{Greyson Davis}
\author[d]{Leonidas A. Moustakas}
\author[c,e]{D. Andrew Howell}
\author[f]{Saurabh W. Jha}
\author[g]{David J. Sand}
\author[h]{Peter Brown}
\author[i]{Ken Shen}

\affil[a]{University of Iowa, Department of Physics \& Astronomy, 203 Van Allen Hall, Iowa City, IA 52242, USA}
\affil[b]{University of Florida, Department of Astronomy, Bryant Space Science Center, Gainesville, FL, USA}
\affil[c]{Las Cumbres Observatory, 6740 Cortona Drive, Suite 102, Goleta, CA 93117-5575, USA}
\affil[d]{Jet Propulsion Laboratory, California Institute of Technology, 4800 Oak Grove Drive, Pasadena, CA 91109, USA}
\affil[e]{Department of Physics, University of California, Santa Barbara, CA 93106-9530, USA}
\affil[f]{Department of Physics and Astronomy, Rutgers, The State University of New Jersey, 136 Frelinghuysen Road, Piscataway, NJ 08854, USA}
\affil[g]{Steward Observatory, University of Arizona, 933 North Cherry Avenue, Tucson, AZ 85721-0065, USA}
\affil[h]{Department of Physics and Astronomy, Texas A\&M University, 4242 TAMU, College Station, TX 77843, USA}
\affil[i]{Department of Astronomy and Theoretical Astrophysics Center, University of California, Berkeley, CA 94720, USA}

\cftpagenumbersoff{figure}
\cftpagenumbersoff{table} 
\begin{document} 
\maketitle

\begin{abstract} 
Our understanding of cosmology is shaped by Type Ia supernovae (SNe Ia), the runaway thermonuclear detonations of white dwarfs via accretion from a companion star. The nature of this companion star is highly debated, with disparate models explaining currently available SNe Ia data. Critical ultraviolet (UV) signatures of SNe Ia progenitors are only observable within the first few days post-detonation. We present the instrument design of UVIa, a proposed SmallSat to make early UV observations of SNe Ia. UVIa conducts simultaneous observations in three photometric channels: far-UV (1500 –- 1800 \AA), near-UV (1800 –- 2400 \AA), and Sloan $u$-band (3000 –- 4200 \AA). UVIa employs two 80 mm double-offset Cassegrain UV telescopes and a similar 50 mm $u$-band telescope, imaging onto three Teledyne e2v CIS120-10-LN CMOS detectors. The UV detectors are delta-doped for enhanced sensitivity, with custom metal-dielectric filters providing further in-band efficiency and red light rejection. The UV optics utilize multi-layer coatings, defining the UV bandpasses and providing additional red light rejection. The instrument design achieves high UV sensitivity (21.5 mag AB) and superior red light rejection ($<$ 10$^{-5}$ throughput), allowing UVIa to make early observations of SNe Ia while serving as a pathfinder for future UV transient telescopes. 
 
\end{abstract}


{\noindent \footnotesize\textbf{*}Fernando Cruz Aguirre,  \linkable{edwinfernando-cruzaguirre@uiowa.edu} }

\begin{spacing}{2}   

\section{Introduction} \label{s:intro} 

Type Ia supernovae (SNe Ia) arise from the thermonuclear runaway of carbon-oxygen core white dwarves (WDs) after mass accretion from a companion star. SNe Ia are scientifically rich astrophysical transient events, providing windows into extreme physical processes and leading to the discovery of the accelerating expansion of our universe via dark energy \cite{1998AJ....116.1009R,1999ApJ...517..565P,2004ApJ...607..665R,2006NuPhA.777..579H,2019MNRAS.482L..70M}. Their status as a cornerstone of modern cosmology comes from the predictable and standardizable nature of their light curves, driven by the decay of $^{56}$Ni into $^{56}$Fe \cite{1993ApJ...413L.105P,1997ApJ...483..565P}. 

We present the optical design and trade study for the Ultraviolet Supernova Ia (UVIa) CubeSat/SmallSat, a dedicated platform designed to capture the early ultraviolet (UV) signatures of SNe Ia. UVIa conducts simultaneous photometry in three bandpasses: far-UV (FUV; 1500 -- 1800 \AA), near-UV (NUV; 1800 -- 2400 \AA), and Sloan $u$-band (3000 -- 4200 \AA). UVIa incorporates several newly available UV technologies into the instrument design, enabling performance comparable to larger UV telescopes. An overview of UVIa's mission concept and science program can be found in Ref. \citenum{HoadleyJATIS}. This manuscript is based on proceedings\cite{2024SPIE13093E..3NC} written for the SPIE Astronomical Telescopes + Instrumentation 2024 conference and provides further details on the UVIa instrument design, UV technologies and performance modeling, and instrument alignment. Section \ref{s:telescopes} describes the optical design of UVIa's telescopes. The UV technologies, which UVIa aims to mature for future NASA missions, are described in Sec. \ref{s:tech_dev}. We present the projected performance of the instrument, described in Sec. \ref{s:performance}. The optical integration and alignment plan is described in Sec. \ref{s:alignment}. 

\section{Double-offset Cassegrain Telescopes} \label{s:telescopes}

The instrument design of UVIa plays a critical role in achieving the goals laid out in UVIa's science program\cite{HoadleyJATIS}. Following from Ref. \citenum{HoadleyJATIS}, the optical design of the instrument must satisfy a few key requirements: 

\begin{itemize}
    \item A co-aligned three-channel system with sensitivity in the FUV, NUV, and $u$-band, to image key progenitor diagnostics in the FUV and NUV, identify distinct explosion pathways via color differences between bands, and calibrate to ground-based $u$-band observations
    \item An angular resolution of $<$ 12'', for resolving SNe Ia from their host galaxies and other contamination sources \cite{2017PASP..129j4502K}
    \item A field of view (FOV) of 1$^{\circ}$$\times$1$^{\circ}$, to simultaneously observe $\geq$ 3 UV-bright calibration stars within a typical SNe Ia field
\end{itemize}

We address these optical design requirements with double-offset Cassegrain telescopes for all three of UVIa's channels. A double-offset Cassegrain telescope consists of an concave, off-axis parabolic (OAP) primary mirror and a convex, off-axis hyperbolic (OAH) secondary mirror. This type of telescope was selected for the several advantages it provides. Compared to a traditional Cassegrain telescope with a primary mirror of equal size, the total collecting area is maximized by offsetting the optical axis of the primary such that the nominal position of the secondary does not obscure the primary. As with a traditional Cassegrain telescope, the figure of the hyperbolic secondary can be selected to either increase or decrease the focal length of the telescope. Double-offset telescopes do come with a number of disadvantages: the spot size at the focal plane increases dramatically with angle of incidence (AoI) when compared to an on-axis system, and optical alignment can become more challenging due to the loss of rotational symmetry \cite{2018OExpr..2624816J}. Additionally, off-axis mirrors are more expensive to fabricate than their on-axis counterparts, especially for certain optical parameters which result in surfaces with extreme curvature.

When designing the double-offset Cassegrain telescopes for UVIa, we utilized an analytic representation of an OAH segment such that the center of the optic has a slope of zero \cite{2023JSynR..30..514G}. The input parameters for this representation are:

\begin{itemize}
    \item $p$, the distance from the OAH segment center to the near focus, the object distance
    \item $q$, the distance from the OAH segment center to the far focus, the image distance
    \item $\theta$, the angle between a plane tangent to the OAH segment center and either the near or far focus, the incidence angle
\end{itemize}

\noindent This representation takes advantage of two properties of hyperbolae: 1) the semi-major axis of a hyperbola, $a$, is related to $p$ and $q$ by $|q-p| = 2a$, and 2) the tangent plane at any given point along a hyperbola bisects the angle between both foci, allowing for a single angle $\theta$ to represent this point. The semi-minor axis $b$ and linear eccentricity $c=\sqrt{a^2+b^2}$ are then calculated as:  

\begin{equation}
\label{eq:OAH_orig_b}
b = \sqrt{pq~sin^2~\theta}
\end{equation}
\begin{equation}
\label{eq:OAH_orig_c}
c = \frac{1}{2}\sqrt{p^2+q^2-2pq~cos~2\theta}
\end{equation}

\noindent To model an OAH in ray tracing programs such as Ansys Zemax OpticStudio, it is useful to convert $a$, $b$, and $c$ of a parent hyperbola into the radius of curvature at the vertex of the parent hyperbola $R_H$ and the conic constant $K_H$ of the parent hyperbola:

\begin{equation}
\label{eq:OAH_orig_R}
R_H = \frac{b^2}{a}
\end{equation}
\begin{equation}
\label{eq:OAH_orig_e}
e_H = \frac{c}{a}
\end{equation}
\begin{equation}
\label{eq:OAH_orig_K}
K_H = -e_H^2
\end{equation}

While initial double-offset Cassegrain designs for UVIa utilized the analytical representations above, we ultimately inverted these equations to define an OAH segment with $R_H$, $K_H$, and $\theta$. Properties of the parent hyperbola, useful for generating plots of the OAH segment in programming languages such as \texttt{python}, are then calculated from these new input parameters as:

\begin{equation}
\label{eq:OAH_invt_a}
a = \frac{-R_H}{1+K_H}
\end{equation}
\begin{equation}
\label{eq:OAH_invt_b}
b = \abs{\sqrt{\frac{-R_H^2}{1+K_H}}}
\end{equation}

\noindent with the AoI $\theta$ determined by the off-axis distance from the parent hyperbola's optical axis. We inverted the input parameters to model OAH secondaries with an equal and opposite radius of curvature to the OAP, $R_P$, in a straightforward manner. By parameterizing the OAH with $R_H$ and $K_H$, equating the radii of curvature of both mirrors is straightforward; doing so results in the Petzval field curvature at the focal plane being zero. Flattening the focal plane yielded significant improvements in spot sizes of off-axis light compared to double-offset designs with OAHs parameterized by $p$ and $q$. 

\begin{figure}[ht!]
    \begin{center}
    \begin{tabular}{c}
    \includegraphics[width=0.98\textwidth]{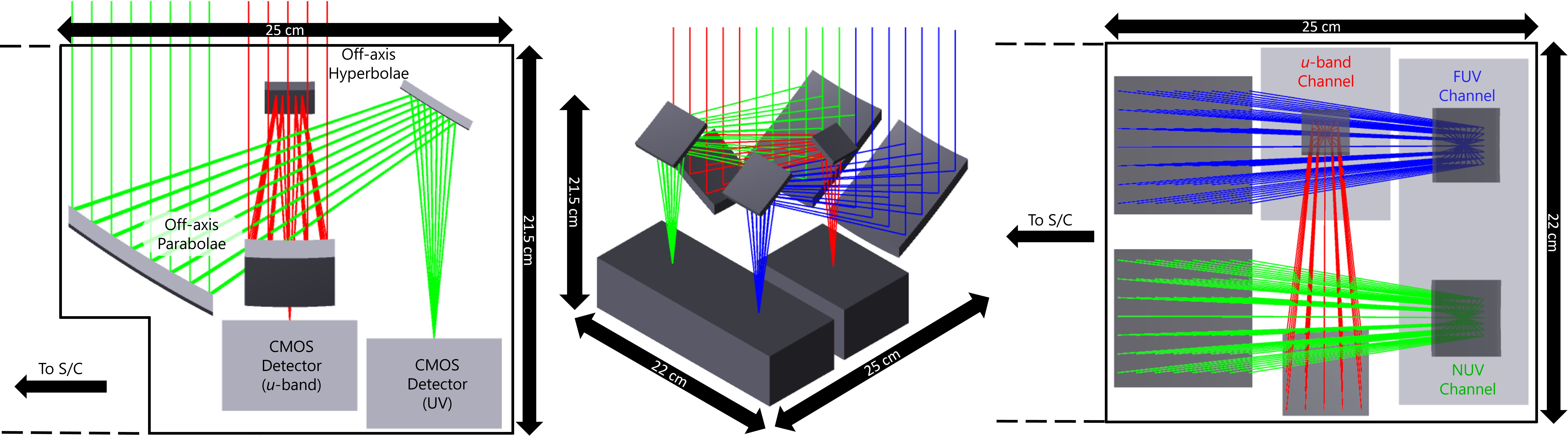}
    \\(a) \hspace{4.5cm} (b) \hspace{4.5cm} (c)
    \end{tabular}
    \end{center}
    \caption{The optical design of UVIa's three photometric channels, as seen from the side (a), an isometric view (b), and from above (c). In views (a) and (b), rays from the target are shown coming from the top of the page.}\label{fig:opto_design} 
\end{figure} 

The optical path of each telescopes is such that the OAP and OAH direct light away by the same angle, similar to a transversal between parallel lines. The off-axis angle of the OAP, $\theta_{OAP}$, is the angle between the optical axis of the OAP towards the focus of the OAP. The center of the OAH is located along the line connecting the OAP center and its focus, with the near-focus of the OAH coincident with the focus of the OAP. The angle $\theta$ used in the analytic representation of the OAH described above is then calculated as $\theta = 180^\circ - \theta_{OAP}$, which deflects light towards the detector along an axis parallel to the optical axis of the OAP. The value of $\theta_{OAP}$ is selected such that the size of the double-offset telescope conforms to the allocated payload volume, while also providing sufficient focal length for meeting the instrument's resolution requirement. We set $R_H$ to be equal and opposite to $R_P$, leaving $K_H$ as the remaining variable of the telescope design. Varying $K_H$ will determine the location of the telescope's focal plane by changing the distance to the OAH's far focus. We select a value of $K_H$ such that there is sufficient room for the channel's detector and supporting electronics below the focal plane. 

We adopt an additional constraint where the offset distance of the OAP is large enough such that the secondary does not obscure the primary. This design approach is utilized to maximize collecting area, subsequently maximizing instrument sensitivity. The offset mirror design also displaces the detectors away from the aperture doors of the instrument. We further increase the collecting area by utilizing square apertures rather than circular apertures. Rectangular apertures have been implemented in the \textit{Colorado Ultraviolet Transit Experiment} (\textit{CUTE}) and the \textit{Supernova Remnants and Proxies for ReIonization Testbed Experiment} (\textit{SPRITE}) instruments to maximize throughput within the rectangular form factor of a CubeSat \cite{2023AJ....165...63F,2023SPIE12678E..06I}. The use of square apertures additionally alleviates alignment challenges associated with off-axis mirrors. 

We design the UVIa telescopes using the framework described above. UVIa was originally designed to fit within the allocated payload volume of a 12U commercial off-the-shelf (COTS) spacecraft bus. This compact instrument design, presented in Fig. \ref{fig:opto_design}, can be adapted to larger CubeSat and SmallSat platforms. The proposed optical design consists of six optics, three metal-oxide-semiconductor (CMOS) detectors (75 mm x 100 mm x 50 mm; described in Sec. \ref{sss:CIS}), and a star tracker (66.0 mm x 56.0 mm x 66.5 mm). When designing the telescopes, the side lengths of the primary mirrors were maximized under the volumetric constraints of a 12U payload. A single optical design is utilized for both the FUV and NUV channels (Sec. \ref{ss:UV}), and another similar design is utilized for the $u$-band channel (Sec. \ref{ss:u-band}). The $u$-band design differs in order to accommodate the third channel within the remaining volume in a COTS CubeSat payload. At the scale of a SmallSat, a single optical design for all three channels is feasible due to the decreased volumetric constraints of the payload. 

\subsection{Ultraviolet Telescopes} \label{ss:UV}

For UVIa's FUV (1500 -- 1800 \AA) and NUV (1800 -- 2400 \AA) channels, we use identical optical designs. The current design has an OAP with an aperture side length of 80 mm, selected to maximize the collecting area of a square aperture while balancing the available payload volume. The OAP redirects light $\sim$63$^{\circ}$ away towards the OAH. The square aperture OAH was sized to fully capture and image light onto the entire active area of the detector, which then defined the instrument FOV. The OAH will redirect incoming light by the same amount as the OAP, $\sim$63$^{\circ}$, directly downward towards UV enhanced CMOS detectors.

While the optical designs of both UV channels are identical, they will each be individually optimized to maximize sensitivity within either the FUV or NUV bandpass. This is accomplished through UV mirror coatings and detector technologies, discussed in more detail in Sec. \ref{ss:optic_tech} and Sec. \ref{ss:cmos_tech}, respectively. These enabling technologies allow for this optical design to meet the instrument requirements for achieving UVIa's science objectives. These technologies also define the bandpasses for each channel. A summary of the UV channel design is given in Tab. \ref{tab:tele_params}.

\subsection{Sloan u-band Telescope} \label{ss:u-band} 

The $u$-band (3000 -- 4200 \AA) channel uses a similar telescope design as the UV telescopes, modified to accommodate the $u$-band channel in the space between the UV primaries and the UV detector (see Fig. \ref{fig:opto_design}). It has a smaller OAP with an aperture side length of 50 mm, sufficient for reaching the desired $u$-band sensitivity requirement. The OAP redirects light $\sim$57$^{\circ}$ away towards the OAH. As with the UV channels, the OAH is sized to fully capture and redirect light onto the full extent of the detector's active area. The $u$-band optics will be coated with Al+MgF$_2$ to enhance in-band reflectivity. The $u$-band channel uses a commercial version of the CMOS detector used for the UV channels, but without UV enhancing technologies. Just before imaging onto the detector, light passes through a Sloan $u$-band filter which defines the bandpass. The channel design is summarized in Tab. \ref{tab:tele_params}.

\begin{table}[h!]
\caption{Design parameters for UVIa's three photometric channels.} \label{tab:tele_params}
\centering
\resizebox{\columnwidth}{!}{
\begin{tabular}{ccccc}
\hline\hline
\textbf{Parameter} & \textbf{UV Primary} & \textbf{UV Secondary} & \textbf{$u$-band Primary} & \textbf{$u$-band Secondary} \\
\hline
Aperture Side Length (mm) & 80.0 & 43.1 & 50.0 & 28.0\\
Radius of Curvature (mm) & -387.7 & 387.7 & -381.0 & 381.0\\
Conic Constant & -1.0 & -20.1 & -1.0 & -20.0\\
Off-axis Distance (mm) & 239.3 & 53.4 & 205.6 & 45.7\\
Off-axis Angle (deg) & 63.37 & 116.63 & 56.72 & 123.28 \\
\hline
\end{tabular}
}
\end{table}

\section{Enabling Technologies} \label{s:tech_dev}

The design of the instrument alone is not sufficient for achieving UVIa's science objectives. We also incorporate several recent technological developments for UV mirrors and detectors, greatly enhancing the instrument's UV sensitivity. Alongside its science program, UVIa is also a technology demonstration platform, and will test these in space for the first time. UVIa directly addresses Tier 1 -- 3 technology gaps in NASA's Astrophysics Technology Gap Priorities\footnote{\url{https://apd440.gsfc.nasa.gov/tech_gap_priorities.html}} by elevating these UV technologies to a Technology Readiness Level (TRL) of 7/8. Future larger-scale UV missions, from Pioneers Class to the next Flagship with UV capabilities, the \textit{Habitable Worlds Observatory} (\textit{HWO}), will benefit from UVIa's TRL maturation program. We divide these technologies below as either Optic Technologies (Sec. \ref{ss:optic_tech}) or Detector Technologies (Sec. \ref{ss:cmos_tech}). A monitoring plan to record the performance of these technologies over the mission lifetime is discussed in Ref. \citenum{HoadleyJATIS}. The optical models presented below were developed using the transfer matrix method (e.g., Ref. \citenum{2010OExpr..1824715T}) and the TFCalc software package (Version 3.5) \cite{tfcalc}. Optical constants for Si, Al, and various metal oxides and metal fluorides were taken from Palik and the Sopra database \cite{tfcalc,1985hocs.book.....P}. Optical constants derived from spectroscopic ellipsometry data, collected using laboratory prepared samples (Horiba UVISEL 2; J. A. Woollam VUV-VASE), were also used in modeling where appropriate.

\subsection{Optic Technologies: Multi-layer Ultraviolet Coatings} \label{ss:optic_tech}

\begin{wrapfigure}{r}{0.50\textwidth}
   \centering
   \includegraphics[width=0.45\textwidth]{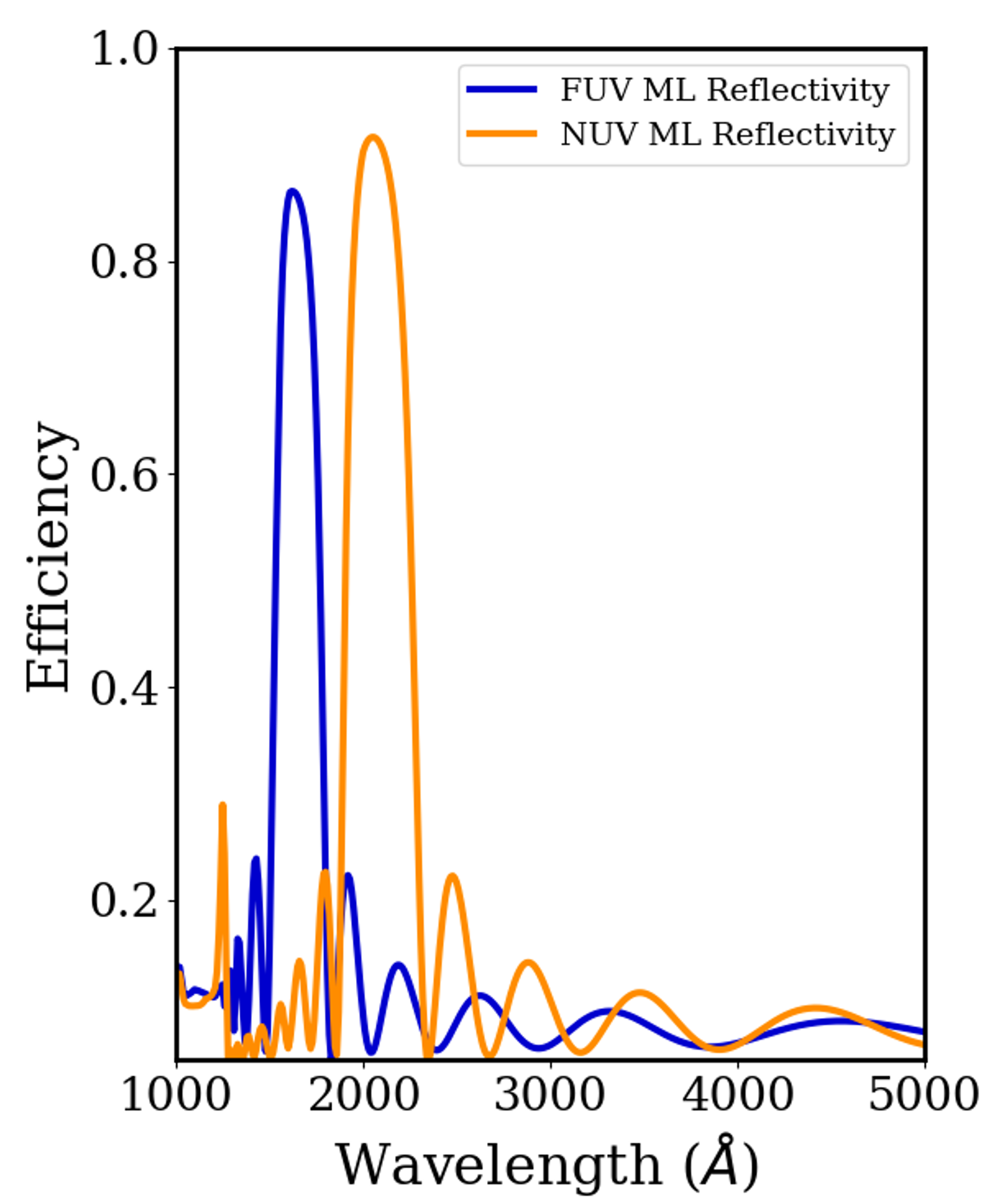}
   \caption{The modeled reflectance of FUV (blue) and NUV (orange) optics with custom ALD ML coatings applied.}\label{fig:ml_coating} 
\end{wrapfigure}

The primary and secondary mirrors of the FUV and NUV channels will utilize multi-layer (ML) dielectric optical coatings. ML coatings produce high reflectance about a central wavelength due to constructive interference between layers. The maximum achievable reflectance of an ML coating is a function of the number of layers in the stack and the optical loss experienced in the constituent materials. Such coatings become challenging in the UV due to the lack of high refractive index materials with sufficiently low loss. A smaller index difference then requires more layers to achieve the same predicted performance. As a result, mechanical effects from residual film stress often limit the ultimate layer count.

Traditional ML coatings in the UV have been produced commercially via physical vapor deposition (PVD) \cite{2005ApJ...619L...1M,2014SoPh..289.2733D}. The Jet Propulsion Laboratory (JPL) is currently developing ML coatings fabricated via atomic layer deposition (ALD) \cite{2017SPIE10401E..19H}. ALD coatings may offer improvements in stability and performance compared to commercial PVD coatings through more precise control of layer thickness; UVIa will test ALD ML coatings in a space environment. 

UVIa's ALD ML coatings are custom developed for each UV channel, defining their bandpasses. Figure \ref{fig:ml_coating} shows the anticipated reflectivities of the baseline ML coatings, which assumes normal incidence, average polarization, and room temperature (295 K). The baseline ML coating design utilizes 8 bilayers (16 total layers) of ALD AlF$_3$ and ALD LaF$_3$ in each of the FUV and NUV channels. Both ALD coating processes utilize a hydrogen fluoride reaction pathway that has been demonstrated at JPL for a variety of metal fluoride thin films \cite{2017SPIE10209E..0PH}. The estimated performance predicted in Fig. \ref{fig:ml_coating} uses refractive index models derived from measurements of individual ALD layers that were characterized by spectroscopic ellipsometry from 1900 -– 9000 \AA, and FUV reflectometry from 1200 -– 2250 \AA. Prototype 16-layer ML coatings of AlF$_3$/LaF$_3$ have been fabricated with a center wavelength of 1400 \AA\ (shorter than the UVIa FUV channel) and demonstrated a peak reflectance of $\sim$90\%. This gives good confidence in the predicted performance of $>$85\% in the FUV channel and $>$90\% in the NUV channel for the baseline design, as the ML reflectance is generally expected to fall off at shorter wavelengths due to intrinsic absorption loss in the constituent layers. The rejection of out-of-band light ($\lambda>3000$ \AA) is projected to be $\leq$10\%. The spacecraft will be kept at an ambient temperature of 295 $\pm$ 20 K; the change in index of refraction in response to a change in temperature is on the order of 10$^{-5}$ -- 10$^{-6}$ K$^{-1}$ for crystal fluoride materials such as the UVIa ML bilayers \cite{1995OptEn..34.1369T,2005SPIE.5904..222L}, and does not significantly impact coating performance.

\subsubsection{Balancing In-Band Reflectivity and Out-of-Band Rejection} \label{sss:bounces}

When designing a UV telescope, it is customary to minimize the number of ``bounces'' due to traditionally low mirror reflectivities ($\sim$60\%)\cite{1994SPIE.2283...12G}. Many advances in the past decade have shown up to 90\% reflectivity \cite{2017ApOpt..56.9941F,2024AAS...24345722Q}. These advanced coatings tend to be broadband UV coatings, which are not well suited for addressing UVIa's instrument requirements. UVIa instead uses ML coatings in its optical design to achieve high in-band reflectance while simultaneously providing out-of-band rejection. Initial designs for the UV telescopes involved using a single prime-focus OAP mirror. This design was abandoned for two reasons, focal length and out-of-band rejection. It was challenging to accommodate a long enough focal length to meet UVIa's resolution requirement within the allocated payload volume of a 12U CubeSat. 

As the mission concept matured, it became apparent that a single ALD ML coated optic could not provide sufficient out-of-band rejection alone; a minimum of two bounces are required to suppress contaminating out-of-band light. This led to a trade study of two mirror systems, which ultimately resulted in the optical design presented in Sec. \ref{ss:UV}. With two mirror bounces in each channel, the effective reflectance is $\geq$70\% in the FUV and $\geq$80\% in the NUV, with $\leq$1\% throughput outside of each bandpass.  

\subsubsection{Angle of Incidence and Coating Performance} \label{sss:aoi}

The baseline ML coatings described above consider a normal AoI, whereas the double-offset Cassegrain telescopes described in Section \ref{s:telescopes} are designed with large off-axis angles of incidence. This is a result of balancing the small payload volume offered by a CubeSat and satisfying science-driven instrument requirements. Up to an AoI of $\sim$25$^\circ$, decreases in performance are negligible and the reflectivity shown in Fig. \ref{fig:ml_coating} can be assumed. As AoI increases, polarization splitting between p-polarized and s-polarized light causes a reduction in overall reflectivity. 

\begin{figure}[ht!]
    \begin{center}
    \begin{tabular}{c}
    \includegraphics[width=0.9\textwidth]{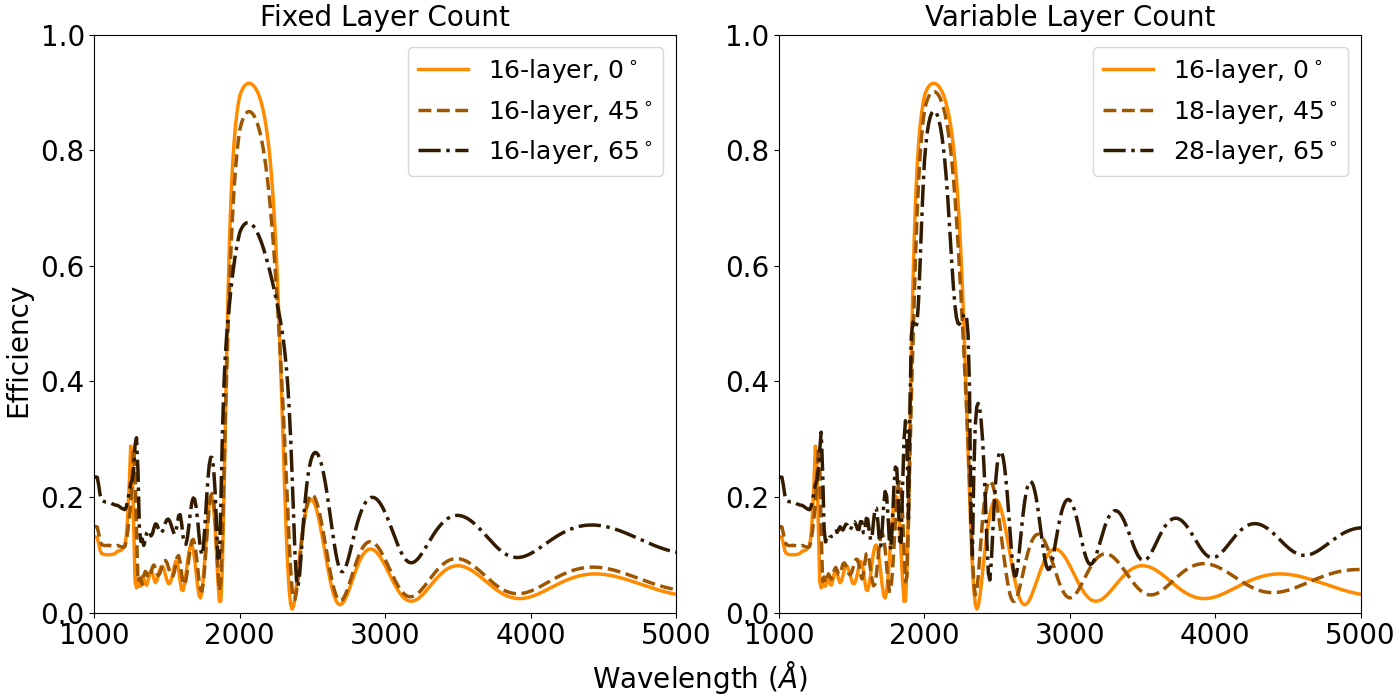}
    \\(a) \hspace{6cm} (b)
    \end{tabular}
    \end{center}
    \caption{Panel (a) shows the reflectivity of the baseline NUV 16-layer ML coating at AoIs of 0$^\circ$ (yellow, solid), 45$^\circ$ (brown, dashed), and 65$^\circ$ (black, dot dashed). Panel (b) shows the anticipated performance of ML coatings with increased layer counts to compensate decreases in performance of the baseline design at larger AoIs.}\label{fig:aoi}
\end{figure}

Figure \ref{fig:aoi}a shows the decrease in in-band reflectivity and out-of-band rejection as AoI increases with all else held constant. By increasing the number of bilayers, in-band reflectivity can be improved at greater AoIs; A total layer count of 28 would provide $\sim$85\% peak reflectivity at an AoI of 65$^\circ$, which is approximately the AoI of the UV telescopes described in Section \ref{ss:UV}, with a factor of $\sim$2 decrease in out-of-band rejection. The performance of this coating is shown in Fig. \ref{fig:aoi}b. The 28-layer model keeps the thickness of each bilayer fixed, and there is room for further optimization by allowing the thickness of each bilayer to vary. UVIa's UV telescopes require a large off-axis angle in order to conform to the available payload volume of a CubeSat; at the scale of a SmallSat, the increased payload volume also increases the design trade space, allowing for smaller AoIs and subsequently less complex coating designs to achieve the baseline ML coating performance.

\subsection{Detector Technologies} \label{ss:cmos_tech}

\subsubsection{CIS-120} \label{sss:CIS}

UVIa employs three Teledyne e2v CIS120-10-LN (``Capella'') CMOS detectors\cite{2021SPIE11852E..0SG} in its optical design, one for each channel. These sensors, when used in rolling shutter (RS) mode, provide low intrinsic noise ($<10$ e$^-$ px$^{-1}$ read noise, $<0.05$ e$^-$ px$^{-1}$ s$^{-1}$ dark current) at an operating temperature of -35$^\circ$C. We baseline the Capella detectors as they are TRL 6 flight-ready COTS devices which satisfy several of UVIa's instrument requirements. In 14-bit RS mode, the detector's anticipated low noise and quantum efficiency (QE) satisfy the sensitivity requirement for the $u$-band. For all three channels, the large 2k $\times$ 2k focal plane satisfies the FOV requirement ($>$1 deg$^2$) and the 10 $\mu$m pixel size satisfies the resolution requirement ($<$6''/px).

The Capella detector comes with built-in processing capabilities, such as integration and stacking of multiple exposures prior to readout. UVIa uses this capability to process and compress stacked science images to be sent to the ground, avoiding additional on-board computing resources which consume power, add mass, and dump heat into the system. During mission operations, passive cooling via the spacecraft's radiators will maintain the nominal operating temperature of -35$^\circ$C. On-chip temperature sensors will monitor and autonomously control device temperatures via heaters installed along the cold chain. Including supporting electronics, each detector is expected to occupy 100 mm $\times$ 75 mm $\times$ 50 mm within the allocated payload volume.

\subsubsection{Delta-doped UV Detectors} \label{sss:DeltaDoped}

While the Capella detector as-is will satisfy the requirements for the $u$-band channel, they are insufficiently UV sensitive for UVIa to achieve its science objectives. For this reason, the two UV detectors will have their in-band QE enhanced further through a now standard delta-doping process performed at JPL \cite{2017JATIS...3c6002N,2018SPIE10709E..0CJ,2023Senso..23.9857H,2024SPIE13093E..04J}. The delta doping process highly concentrates a dopant into a very thin layer, usually one-to-two atomic layers in a crystal plane, to enable the detection of high-energy (e.g., X-ray, UV) photons. This is achieved by eliminating a ``dead'' layer on the back surface of a silicon sensor where these photons are absorbed without detection. 

UVIa's UV Capella detectors will be back-illuminated and optimized for enhanced UV sensitivity via delta-doping. These same processes have been utilized previously for charge-coupled device (CCD) sensors on the SPARCS \cite{2024SPIE13093E..04J} and FIREBall-2 \cite{2020JATIS...6a1007K} missions, and have demonstrated $\sim$100\% internal QE \cite{2012ApOpt..51..365N}. Protoype delta-doped Capella detectors are currently being characterized at JPL, with a path towards TRL 5 already underway.

\subsubsection{Metal-dielectric Filters} \label{sss:MDF}

Delta-doped silicon detectors remain sensitive to optical light. SNe Ia are optically bright, potentially $\sim$100$\times$ brighter in the optical band than in the UV (e.g., Ref. \citenum{2014MNRAS.439.1959M}). Emerging metal-dielectric filter (MDF) technology provides the required optical-light suppression ($\lambda>3000$ \AA) \cite{2015ApOpt..54.3507H,2017SPIE10209E..0PH,2021SPIE11821E..1AH}. Metal-dielectric coatings can be layered to act as UV bandpass filter. These filters have been used in previous UV instruments as standalone elements \cite{1990ASPC....8..217T,2005SSRv..120...95R}. By integrating these layers directly on silicon detectors, in-band transmission and out-of-band rejection can be improved \cite{2015ApOpt..54.3507H,2017SPIE10209E..0PH}. MDF provides waveband selection control, defined by the number of MDF layers laid on the device.

\begin{wrapfigure}{r}{0.50\textwidth}
   \centering
   \includegraphics[width=0.45\textwidth]{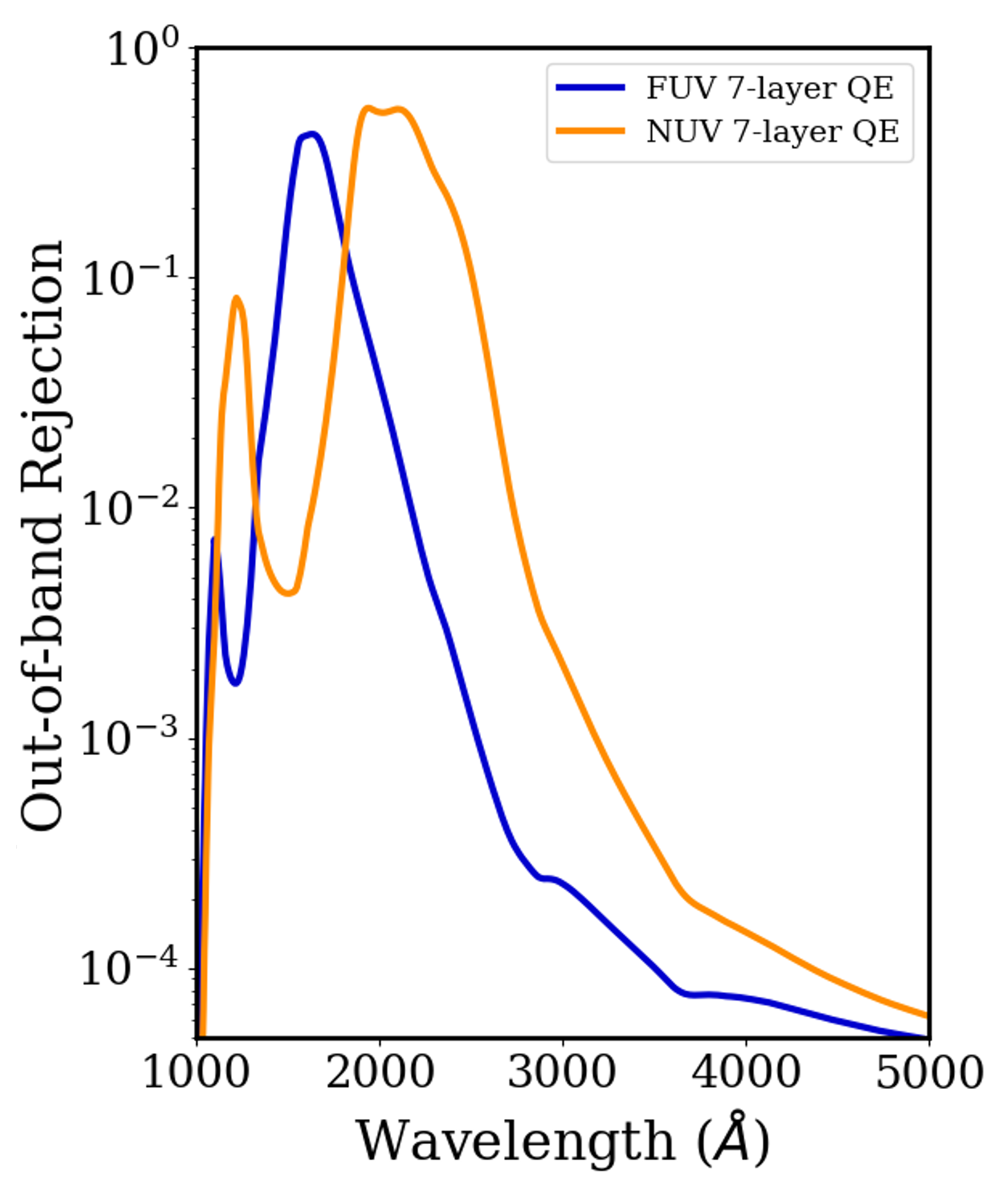}
   \caption{The out-of-band rejection of UVIa's FUV (blue) and NUV (orange) detectors. In-band enhancement is achieved via delta-doping, and red-light rejection via MDF coatings.}\label{fig:mdf_coating}
\end{wrapfigure} 

UVIa's MDF coatings will be custom tailored for each UV channel by JPL. The MDF designs utilize multilayer combinations of ALD AlF$_3$ and thermally evaporated Al. The 7-layer FUV filter, centered about 1600 \AA, is similar to a design recently implemented on CCDs for the SPARCS CubeSat mission\cite{2024SPIE13093E..04J}, as well as designs implemented on CMOS image sensors for the upcoming UVEX mission \cite{2022JATIS...8b6004G}. The 7-layer NUV design is centered about 2100 \AA. In conjunction with delta-doping, the FUV channel provides a peak QE performance of $\sim$40\% and the NUV channel provides $\sim$55\% peak QE. The sensitivity and out-of-band rejection attainable in the UV channels via the Detector Technologies are illustrated in Fig. \ref{fig:mdf_coating}. UVIa's double-offset Cassegrain telescopes are designed to direct on-axis light at a normal incidence to the CMOS sensors. Off-axis light will, at most, reach the sensors at an AoI of $\sim$5$^\circ$ with the current CubeSat design. Similar to the ML coating performance at low AoIs, the MDF filters show negligible decreases in performance at an AoI of 5$^\circ$.

\section{Projected Instrument Performance} \label{s:performance}

\begin{figure}[ht!]
    \begin{center}
    \begin{tabular}{c}
    \includegraphics[width=0.8\textwidth]{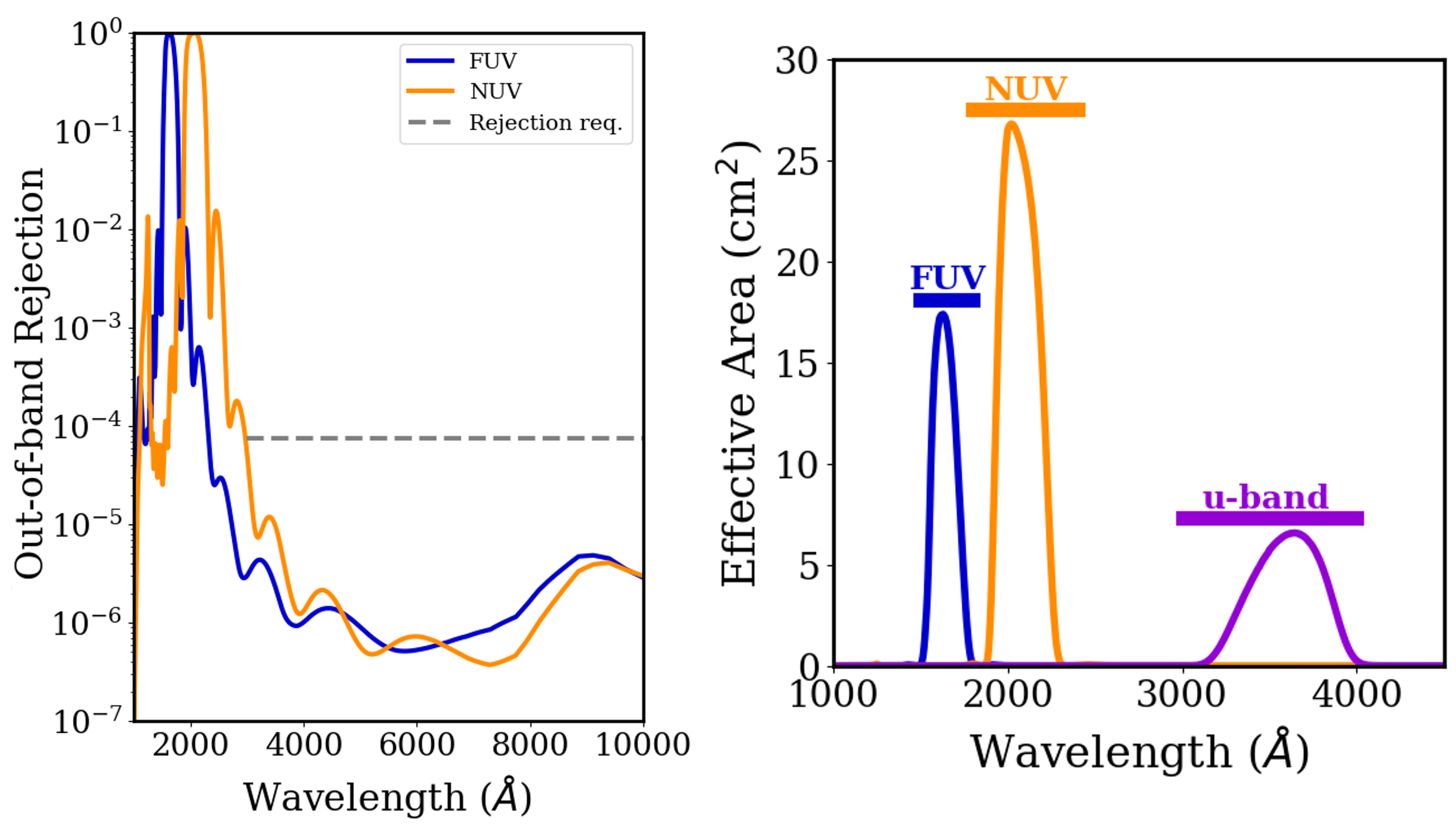}
    \\(a) \hspace{6cm} (b)
    \end{tabular}
    \end{center}
    \caption{The out-of-band rejection in the UV channels (a) and the effective area in all three channels (b).}\label{fig:rRej_aEff} 
\end{figure}

Altogether, the optical design and the technologies described in Sec. \ref{s:telescopes} and Sec. \ref{s:tech_dev}, respectively, exceed the instrument requirements. Table \ref{tab:inst_perf} compares science-driven instrument requirements to the projected performance of the UV and $u$-band channels, and then lists the margin or multiplier against the requirement. Not listed in Tab. \ref{tab:inst_perf} is the signal-to-noise ratio (SNR) and out-of-band rejection; see Ref. \citenum{HoadleyJATIS} for a discussion on these instrument requirements. Here, we note that the projected SNR performance results from the culmination of UVIa's optical design and all state-of-the-art UV technologies described above. The instrument's out-of-band rejection in the two UV bands is shown in Fig. \ref{fig:rRej_aEff}a, and the effective area of all three bands are shown in Fig. \ref{fig:rRej_aEff}b.

\begin{table}[ht!]
\caption{The instrument requirements and projected performance of UVIa.} \label{tab:inst_perf}
\centering
\resizebox{0.8\columnwidth}{!}{
\begin{tabular}{cccc}
\hline\hline
\textbf{Parameter} & \textbf{Requirement} & \textbf{Performance} & \makecell{\textbf{Margin or}\\\textbf{Multiplier}} \\
\hline
\makecell{Bandpass\\(\AA)} & \makecell{FUV: 1500 -- 1800\\NUV: 1800 -- 2400\\$u$-band: 3000 -- 4200} & \makecell{FUV: 1500 -- 1800\\NUV: 1800 -- 2400\\$u$-band: 3000 -- 4200} & --- \\ \hline
\makecell{Angular\\Resolution ('')} & $<$ 12 & \makecell{UV: 7.1\\$u$-band: 7.5} & \makecell{UV: 69\%\\$u$-band: 61\%} \\ \hline
FOV (deg$^2$) & $>$ 1 & 6.25 & 6.25$\times$ \\ \hline
\hline
\end{tabular}
}
\end{table}

The on-axis point spread function (PSF) of the UV channels are within 1'', and the PSF of the $u$-band channel is projected to be 2.4''. At present, a simple analysis was performed to estimate the degradation of angular resolution caused by thermal defocusing of UVIa's three telescopes. Amongst UVIa's mission requirements is that the spacecraft will vary no more than $\pm$20 K about an ambient temperature of 295 K. A thermal multi-configuration was set up in OpticStudio using the thermal properties for fused silica for the UVIa optics. The thermal expansion or contraction of the mirror substrate was estimated within the above temperature range, and the spot size at the detector was then measured. From this analysis, the spot sizes of the UV channels increase by $\sim$60 -- 70\%, whereas the spot size of the $u$-band telescope increase by $\sim$8\%. Spacecraft jitter, however, is the primary cause for degradation. Jitter is expected to circularize the square on-axis PSF. Accounting for the expected jitter of a COTS CubeSat/SmallSat spacecraft, UVIa's spot sizes in each channel are projected to be $\lesssim$ 7.5'' for incident light which is on-axis. The $<$12'' requirement is satisfied for incident light up to a few arcminutes off-axis. Table \ref{tab:opt_bud} summarizes UVIa's optical budget for on-axis light, appropriate for the imaging of SNe Ia.

\begin{table}[h!]
\caption{The optical error budget for UVIa's expected performance.} \label{tab:opt_bud}
\centering
\begin{tabular}{cccc}
\hline\hline
\textbf{Parameter} & \textbf{FUV} & \textbf{NUV} & \textbf{$u$-band}\\
\hline
\makecell{Required Spatial Res.\\(arcsec)} & 12.0 & 12.0 & 12.0 \\
\makecell{On-axis PSF FWHM*\\(arcsec)} & 0.7 & 0.8 & 2.4 \\
\makecell{Spacecraft Jitter + Pointing\\(arcsec RMS)} & 7.0 & 7.0 & 7.0 \\
\makecell{Thermal Defocus*\\(arcsec RMS)} & 1.0 & 1.0 & 1.0 \\
\makecell{\textbf{Projected Spatial Res.}\\(arcsec)} & \textbf{7.1} & \textbf{7.1} & \textbf{7.5} \\
\textbf{Margin Against Required} & \textbf{69\%} & \textbf{69\%} & \textbf{61\%} \\
\hline
\end{tabular}
\begin{center}
    *Includes 25\% margin against projected performance
\end{center}
\end{table}

The FOV requirement for UVIa is driven by having at least three calibration stars present within any given SNe Ia field, whereas the resolution requirement is driven by resolving SNe Ia from contaminating sources, e.g., their host galaxies. The double-offset Cassegrain telescopes, limited by spacecraft jitter, exceed the angular resolution requirement by $\sim$65\% in each channel. The large detector size is capable of imaging an FOV of 2.5$^\circ$ $\times$ 2.5$^\circ$ (6.25 deg$^2$) in each channel. At a factor of 6.25$\times$ larger than the requirement, more calibration stars can be captured during observations of SNe Ia. 

\subsection{The Double-Offset Field of View} \label{ss:inst_alt}

\begin{figure}[ht!]
    \begin{center}
    \begin{tabular}{c}
    \includegraphics[width=0.9\textwidth]{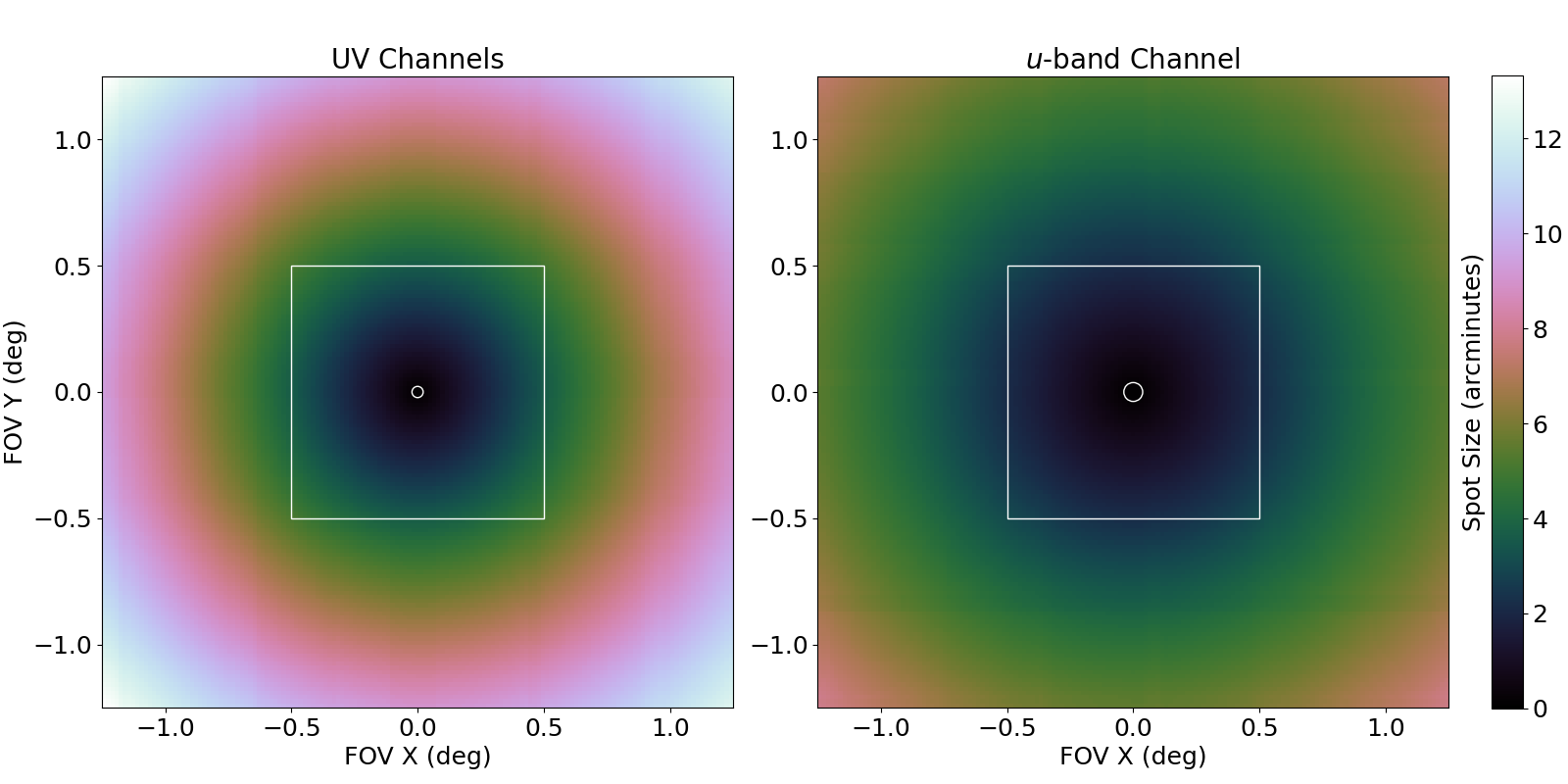}
    \\(a) \hspace{6.25cm} (b)
    \end{tabular}
    \end{center}
    \caption{The PSF field maps of the UV channels (a) and the $u$-band channel (b). The white square represents the 1 deg$^2$ FOV requirement, and the white circle encloses the region where the $<$12'' spatial resolution requirement is satisfied.}\label{fig:field_map} 
\end{figure}

The current optical design presented in Sec. \ref{s:telescopes}, designed for a COTS CubeSat payload, will sufficiently resolve SNe Ia when placed on-axis while simultaneously capturing $\geq 3$ calibration stars off-axis. The spatial resolution for off-axis light is considerably degraded, as shown in Fig. \ref{fig:field_map}, due to the large $\sim$65$^\circ$ off-axis angle of the double-offset system. This angle is necessary to balance the required plate scale at the focal plane and the volumetric constraints of a CubeSat payload. UVIa's $<$12'' spatial resolution requirement is satisfied within the central region of the FOV in each channel (encircled in white in Fig. \ref{fig:field_map}), requiring SNe Ia to be positioned within $\sim$1.3' from FOV center. COTS CubeSat and SmallSat buses are capable of satisfying this pointing requirement. The projected spot sizes near the edges of the instrument FOV exceed $\sim$10' in the UV and $\sim$5' in the $u$-band. Near the edges of the central 1 deg$^2$ FOV region, the projected 3 -- 5' UV spatial resolution approaches the general separation between UV-bright targets in the sky \cite{2017ApJS..230...24B}. For some SNe Ia fields, UV-bright calibration stars may not be resolved towards the edges of the central 1 deg$^2$ FOV. UVIa's success does not rely on resolving individual calibration stars, rather it relies on 1) resolving SNe Ia positioned at FOV center, and 2) identifying at least three calibration stars elsewhere in the FOV. 

The FOV performance of UVIa's double-offset Cassegrain telescopes may not be ideal for other astronomical applications, requiring alterations for reducing degradation. The off-axis performance of these telescopes depends heavily on the off-axis angle of the system, determined by $\theta_{OAP}$: a lower off-axis angle will significantly improve degradation away from FOV center. If there is a constraint which requires a large off-axis angle, increasing the focal length of the system will also improve FOV performance. UVIa's volumetric constrains prevented both lower off-axis angles and greater focal lengths on a COTS CubeSat platform; the increased volume provided by a SmallSat platform increases the trade space for double-offset designs with lower off-axis angles, greater focal lengths, or both. Other applications with differing science requirements, and thus differing optic and detector footprints, may also find more optimal solutions for these applications within a CubeSat payload. 

\section{Optical Integration and Alignment} \label{s:alignment}

While CubeSats and SmallSats provide relatively inexpensive access to space, their compact form factors place constraints on the focal length of a telescope. UVIa's double-offset Cassegrain telescopes are relatively fast: $f/3.9$ for the UV telescopes, and  $f/5.7$ for the $u$-band telescope. Faster telescopes generally require tighter alignment tolerances in order to perform as designed, and this is exacerbated when introducing off-axis optics. UVIa additionally requires all three of its telescopes to be co-aligned with each other. For these reasons, we anticipate there to be challenging alignment tolerances which must be met to achieve the projected performance of the instrument.

The optical alignment plan for UVIa takes inspiration from that of the \textit{Aspera} SmallSat\cite{2021SPIE11819E..03C,2024arXiv240715391K}, due to similarities in optical design (fast telescopes, off-axis optics, two mirror + detector) and associated alignment challenges. A computer generated hologram (CGH) can be used for optical alignment in a quick and inexpensive manner \cite{2022SPIE12213E..0KS,2024SPIE13093E..2VA}, and has been demonstrated on the two mirror focal corrector on board FIREBall-2  \cite{2024SPIE13093E..2VA}. The optical alignment approach for \textit{Aspera}, detailed in Ref. \citenum{2024arXiv240715391K} uses a 3D scanner, an interferometer, a custom CGH, and optical mounts which allow for adjustments to the telescope optics in 6 degrees of freedom (DoF). Detector alignment in Ref. \citenum{2024arXiv240715391K} is conducted in a vacuum chamber with a 3 DoF mount for the detector for alignments and collimated UV beam. 

The optical design of UVIa makes it possible to follow a similar alignment approach. A custom CGH will be fabricated for UVIa's optical design, and will be mounted onto the payload's optical bench. An optic will first be placed into its mount, and the bench will be scanned with a 3D scanner. The alignment of the optic is verified by comparing the scan to a CAD model, and shims will be used to make coarse adjustments. An interferometer will illuminate the optic(s) in each channel with a collimated beam, and the CGH will reflect a portion of the beam back to the interferometer. Based on the interference pattern, precision adjustments can be made with actuators on the optic's mounts. The OAPs will be mounted and aligned, with the CGH ensuring co-alignment of the channels. The OAHs will then be mounted and aligned. The telescope will then be staked, and a final verification of telescope focus will be made prior to removing the CGH. These alignment procedures will be conducted at the anticipated ambient temperature of the spacecraft (295 K), and the instrument spot size throughout a thermal range of 275 - 315 K will be measured. 

To align the Capella detectors, coarse alignment can be performed at ambient pressure. Precision alignment will require a collimated UV light source, which further requires the instrument to be aligned within a vacuum chamber. A large ($\sim$500 mm) UV collimator is planned to be assembled ahead of UVIa to support the integration and calibration of the instrument. Vacuum alignment will be performed by illuminating all three channels, and using vacuum-safe actuators to position the Capella detectors at the telescope focus. The detectors will then be staked and a final optical verification under vacuum will be performed. Vibration and thermal testing will be conducted to verify system alignment under the mechanical and thermal stress during launch. Metering structures will be designed and incorporated within each channel to maintain optomechanical alignment in orbit. 

\section{Summary} \label{s:summary}

UVIa was designed to conduct a compelling science program, shedding light on the origins of SNe Ia. To carry this program out, three co-aligned double-offset Cassegrain telescopes were designed to image SNe Ia in the FUV (1500 -- 1800 \AA), NUV (1800 -- 2400 \AA), and Sloan $u$-band (3000 -- 4200 \AA). The $f/3.9$ FUV and NUV telescopes utilize 80 mm square apertures, and the $f/5.7$ $u$-band channel utilizes a 50 mm square aperture. All primary apertures are unobscured by their secondaries, maximizing the effective area within each channel. The UVIa telescopes fit within typical volumes allocated to CubeSat and SmallSat payloads, while also satisfying the angular resolution requirement for resolving SNe Ia from their host galaxies ($<$ 12'') and providing a large enough FOV to simultaneously image $\geq$ 3 calibration stars in each channel ($>$ 1 deg$^2$). 

Several novel UV technologies are employed by UVIa. ML dielectric optical coatings will be designed and applied to UVIa's FUV and NUV mirrors, defining each channel's bandpass. The baseline, 16-layer, normal incidence coatings provide $>$ 85\% reflectivity in the FUV, and $>$ 90\% reflectivity in the NUV, with similar anticipated performance at greater AoIs with increased layer counts. Altogether, the ML coated mirrors provide $\leq$1\% out-of-band throughput. Each channel is equipped with a Capella CMOS sensor, providing low intrinsic noise and built-in processing capabilities. The UV CMOS sensors will undergo a standard delta-doping procedure to enhance in-band QE. FUV and NUV 7-layer MDF coatings will be applied directly to the UV CMOS sensors to further suppress out-of-band light. These UV technologies, listed as Tier 1 -- 3 Technology Gap Priorities, will be elevated to TRL 7/8 in preparation for future UV-capable missions such as \textit{HWO}.

The double-offset Cassegrain telescopes provide ample margin on spatial resolution (61 -- 69\%) and FOV (6.25$\times$) over the instrument requirements. The fast and off-axis nature of these telescopes, however, requires tight tolerances to be met in order to deliver this performance. UVIa's telescope optics and detectors will be aligned through a procedure developed through the \textit{Aspera} SmallSat and FIREBall-2 balloon missions \cite{2024SPIE13093E..2VA,2024arXiv240715391K}. Through its optical design, enabling UV technologies, and alignment plan, UVIa will be robust to unanticipated degradation in performance.


\subsection*{Disclosures}
The authors declare that there are no financial interests, commercial affiliations, or other potential conflicts of interest that could have influenced the objectivity of this research or the writing of this paper.


\subsection* {Code, Data, and Materials Availability} 
There is no associated code or supporting data for this paper.



\subsection* {Acknowledgments}
Part of this research was carried out at the Jet Propulsion Laboratory, California Institute of Technology, under a contract with the National Aeronautics and Space Administration (80NM0018D0004).

Time-domain research by the University of Arizona team and D.J.S.\ is supported by National Science Foundation (NSF) grants 2108032, 2308181, 2407566, and 2432036 and the Heising-Simons Foundation under grant \#2020-1864.


\bibliography{refs} 
\bibliographystyle{spiejour}   


\vspace{2ex}\noindent\textbf{Fernando Cruz Aguirre} is a Postdoctoral Research Scholar at the University of Iowa. Fernando received his Ph.D. from the University of Colorado, Boulder, where his thesis focused on the FUV radiation environments of exoplanet hosts. He was trained in UV instrumentation through the SISTINE sounding rocket. Fernando was selected for the inaugural Astrophysics Mission Design School, developing a mock UV Astrophysics Probe concept. Currently, Fernando fabricates and characterizes UV gratings, and is involved with several UV missions. 


\vspace{1ex}
\noindent Biographies and photographs of the other authors are not available.

\listoffigures
\listoftables

\end{spacing}
\end{document}